\documentclass[amssymb,amsmath,pra,showpacs,twocolumn,floatfix,superscriptaddress]{revtex4}
\usepackage{graphicx}
\usepackage{lastpage}
\usepackage{amssymb}

\begin{document}

\title{Tunable photon-assisted bleaching of three-level systems for noise filtration}

\author{Chun-Hsu Su} \email{chsu@ph.unimelb.edu.au}
\affiliation{Centre for Quantum Computer Technology, School of Physics, The University of Melbourne, Victoria 3010, Australia}

\author{Andrew D. Greentree}
\affiliation{School of Physics, The University of Melbourne, Victoria 3010, Australia}

\author{Raymond G. Beausoleil}
\affiliation{Information and Quantum Systems Lab, Hewlett-Packard Laboratories, 1501 Page Mill Road, MS1123, Palo Alto, California 94304, USA}

\author{Lloyd C. L. Hollenberg}
\affiliation{Centre for Quantum Computer Technology, School of Physics, The University of Melbourne, Victoria 3010, Australia}

\author{William J. Munro}
\affiliation{NTT Basic Research Laboratories, NTT Corporation, 3-1 Morinosato-Wakamiya, Atsugi-shi, Kanagawa-ken 243-0198, Japan}

\author{Kae Nemoto}
\affiliation{National Institute of Informatics, 2-1-2 Hitotsubashi, Chiyoda-ku, Tokyo 101-8430, Japan}

\author{Timothy P. Spiller}
\affiliation{Quantum Information Science, School of Physics and Astronomy, University of Leeds, Leeds LS2 9JT, UK}

\date{\today}

\begin{abstract}
Electromagnetically-induced transparency (EIT) exploits quantum coherence to burn subnatural linewidth holes within a spectral line. We investigate the less explored properties of EIT to effect absorptive nonlinear processes without restrictions on the relative intensities of pump and probe fields. We show that a three-level medium under imperfect EIT conditions can generate a form of bleaching that is qualitatively similar to two-state saturable absorption. This scheme has the advantages of greater sensitivity to signal intensity and controllability over its bleaching intensity level post-fabrication. Such effects could prove useful for noise filtration at very low light levels.
\end{abstract}

\pacs{32.30.Jc, 42.50.Gy, 42.65.An}

\maketitle

\section{Introduction}
An ability to control systems down to the limits imposed by quantum mechanics is expected by many to usher a new era in technology, the so-called `second quantum revolution'~\cite{dowling03}. Although often one is interested in harnessing effects not possible, or not practical to achieve with classical physics, quantum processes can also be used to enhance classical processes, and often provide limiting cases in terms of fidelity or power for classical effects. One of the most important and well-studied fields showing quantum control is nonlinear optics. Coherently prepared atoms or atomic systems, often under conditions of electromagnetically-induced transparency (EIT), can be shown to exhibit large, lossless nonlinearities~\cite{harris90} that can be useful for quantum logic~\cite{harris98,ottaviani06,greentree09}. Here we turn our attention to a nonlinearity that is less explored in the quantum space, namely absorptive nonlinearity.

Saturable absorption is a well-known nonlinear process. An optical field impinges on a medium, usually treated as an ensemble of two-state systems~\cite{hercher67,atkinson94}. For low intensities, the field is greatly attenuated, but at high fields, as the medium is saturated, the absorption is relatively low. This effect is useful for many practical tasks, including analog-to-digital conversion~\cite{valley07}, passive mode-locking of laser radiation~\cite{steinmeyer06} and sub-diffraction imaging in confocal fluorescence microscopy~\cite{fujita07}.

\begin{figure}[tb!]
\centering
\includegraphics[width=0.9\columnwidth,clip]{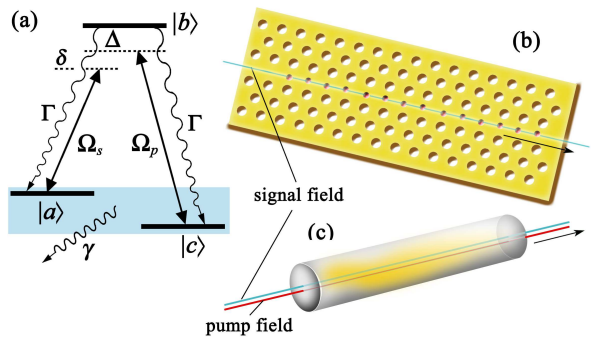}
\caption{(a) Three-level $\Lambda$-type atoms as a nonlinear filter whose bleaching intensity level can be tuned with the pump field. The signal (pump) field couples to the $|a\rangle-|b\rangle$ ($|b\rangle-|c\rangle$) transition with coupling rate $\Omega_s$ ($\Omega_p$). The excited state $|b\rangle$ decays to the pair of ground or metastable states at a total rate $2\Gamma$. The low-frequency coherence dissipates at a dephasing rate $\gamma$, which is responsible for the absorption in the EIT window. (b) Proposed configuration in solid-state environment: atomic systems along a single-mode waveguide constructed from a photonic-bandgap lattice, interacting with the signal beam and an uniform pump field. (c) An alternative implementation with copropagating fields using atomic vapour cell.}
\label{fig:schematics}
\end{figure}

In this work, we analyse EIT-based absorptive nonlinearities in detail. Usually EIT is used as a means of generating a medium with zero absorption and high dispersion, which can be the enabler of large lossless optical nonlinearities~\cite{harris90}. EIT employs two coherent fields (e.g., signal and pump) driving a Raman transition between two, long-lived ground states via a shared excited state, in the $\Lambda$ configuration, as depicted in Fig.~\ref{fig:schematics}(a). When the ground-state decoherence is absent, quantum interference is complete and a long-lived dark state is observed, which manifests ideally as a perfect transparency window at the two-photon resonance point. The associated steep dispersion has recently been shown to allow sub-nanometer-scale resolution for microscopy, surpassing the the two-level system based technique~\cite{kapale10}. 

Ground-state dephasing plays a critical role in practical EIT schemes. In general, the coherence responsible for EIT must compete with decoherence, and this ultimately limits the depth and width of the transparency window. However, this competition allows for saturable absorption. Kocharovskaya \textit{et al.} have shown that the absorptive nonlinearity can occur even when the radiation intensity is much less than that needed for saturation of the transition~\cite{koch86,khanin90,koch90}. The effect is termed coherent bleaching.

Here we investigate this bleaching effect under conditions that are less restrictive than those considered previously in Refs.~\cite{koch86,khanin90,koch90}. In particular we show that the three-state system provides increased flexibility over the onset of the nonlinearity with the signal absorption tunable with the pump field post fabrication, typically absent in two-state schemes. This is distinct from saturation and coherent bleaching because photon-assisted nonlinearity occurs even when both the coherent and saturation nonlinearities are weak. Moreoever the absorption scaling with intensity can be quadratic under certain conditions, and thus can provide a greater sensitivity for applications such as noise filtration and analog-to-digital conversion applications. Given that our focus is on this enhanced absorption properties, this work is also distinct from the effect of tunable electromagnetically-induced absorption in doubly dressed two-level systems~\cite{ian10}. 

This paper is organized as follows: In Sec.~\ref{sec:preliminary} we briefly review EIT and describe the absorptive nonlinearities under various strength of spontaneous decay, dephasing and coherent couplings in the steady state. In Sec.~\ref{sec:opticallythick} we consider propagating fields in optically-thick media in two possible signal-pump arrangements shown in Fig.~\ref{fig:schematics}(b,c). We also explore their application for noise filtration. Finally in Sec.~\ref{sec:impl} we briefly explore some possible implementations of our scheme.

\section{Optical response}
\label{sec:preliminary}
The essential features of EIT and its application to modifying the linear and nonlinear optical properties of a light field can be quantitatively described using a semiclassical analysis~\cite{beausoleil04}. We consider a dilute ensemble of identical $\Lambda$-type atomlike systems. Our model $\Lambda$-type atom is depicted schematically in Fig.~\ref{fig:schematics}(a). Transitions between the excited state $|b\rangle$ and the two ground states $|a\rangle$ and $|c\rangle$ are dipole allowed. A near-resonant pump field of frequency $\omega_p$ is applied on the $|c\rangle-|b\rangle$ transition with detuning $\Delta = (\omega_{bc}-\omega_p)$ and Rabi frequency $\Omega_p$. The $|a\rangle-|b\rangle$ transition is driven by the signal field, with frequency $\omega_s$, detuning $(\Delta+\delta)= (\omega_{ab}-\omega_s)$ and Rabi frequency $\Omega_s$, where $\delta$ is the two-photon detuning between the low energy states. We study the absorption of the signal field ($\Omega_s$) (\textit{e.g.,} for signal processing applications) and allow the signal to be more intense than the coupling field ($\Omega_p$). Hence our analysis is similar in spirit to the work on strong probe EIT \cite{wielandy98}, parametric EIT \cite{muller00}, and the dispersive regime for quantum gates \cite{greentree09}.

In the rotating-wave approximation, the coherent dynamics of the semiclassical atom-field system are described by the Hamiltonian,
\begin{equation}
	\mathcal{H}/\hbar = \Delta\sigma_{bb} + \delta\sigma_{aa} + (\Omega_p\sigma_{cb} + \Omega_s\sigma_{ab} + {\rm h.c.}),
	\label{eq:Ham}
\end{equation}
where $\sigma_{ij} = |i\rangle\langle j|$ are the atomic projection operators. The evolution equation of the atomic density operator $\rho$ of the system is,
\begin{equation}
	 \frac{d\rho}{dt} = -\frac{i}{\hbar} [\mathcal{H},\rho] + \Gamma\Big(\mathcal{L}[\sigma_{ab}] + \mathcal{L}[\sigma_{cb}]\Big) + \gamma\mathcal{L}[\sigma_{aa}-\sigma_{cc}],
	 \label{eq:master}
\end{equation}
where $\Gamma$ and $\gamma$ are the rate of spontaneous emission from the excited state and the dephasing rate of the ground states respectively, and the standard Liouvillian is used
\begin{equation}
	\mathcal{L}[B] = B\rho B^\dagger - \frac{1}{2}\Big(B^\dagger B \rho + \rho B^\dagger B\Big),
\end{equation}
where $B$ is some operator (not necessarily Hermitian) describing the loss channel.  The signal field in the medium responds to (the expectation value of) the polarization generated by the applied fields or, equivalently the linear optical susceptibility $\chi$. In terms of the off-diagonal matrix elements $\rho_{ij} = \langle i|\rho|j\rangle$ per atom, the polarization in the direction of the dipole moment $d_{ab}$ is~\cite{beausoleil04}
\begin{equation}
	P = \frac{1}{2}\epsilon_r \epsilon_0\chi_{ab} E_s + {\rm h.c.} = \langle d \rangle =  \mathcal{N}\Big(\rho_{ab}d_{ab} + {\rm h.c.}\Big),
	\label{eq:polar}
\end{equation}
where $E_s$ is the classical electric field amplitude of the signal, $\epsilon_0$ and $\epsilon_r$ are the permittivity of free space and the relative permittivity respectively, and $\mathcal{N}$ is the atomic density. The atom-field interactions are taken to be electric dipole in nature so that the Rabi frequencies $\Omega_s = d_{ab}E_s/\hbar$ and $\Omega_p = d_{bc}E_p/\hbar$, where $E_p$ is the field amplitude of the pump, $d_{ij}$ is the dipole moment of the $|i\rangle\rightarrow |j\rangle$ transition and it is related to the radiative decay rate by $d_{ij}^2 = 3\pi\hbar\epsilon_0 \epsilon_r c^3\Gamma/(\omega_{ij}^3 \eta)$ in the medium with refractive index $\eta$. Following Eq.~\ref{eq:polar}, the signal field sees the first-order complex susceptibility
\begin{equation}
	\chi_{ab} = \frac{2\mathcal{N} d_{ab}^2}{\hbar \epsilon_r \epsilon_0}\frac{\rho_{ab}}{\Omega_s},
	\label{eq:chi}
\end{equation}
and is therefore attenuated according to the absorption coefficient in the units of inverse length,
\begin{equation}
	\alpha = \frac{\omega_s}{n c}{\rm Im}[\chi_{ab}],
	\label{eq:coeffdef}
\end{equation}
where $n$ is the bulk refractive index. It is related to the real part of the susceptibility by $n^2 = 1 + {\rm Re}[\chi_{ab}]$, and contributes to dispersions and phase shift.

Before proceeding to investigate the absorption nonlinearity of the three-state system, it is instructive to briefly review the well-known result of the two-state absorber. The corresponding absorption coefficient can be derived using the above prescription with a two-state Hamiltonian $\mathcal{H}/\hbar = \delta\sigma_{bb} + (\Omega_s\sigma_{ab} + {\rm h.c.})$. We calculate the steady-state solution of the density matrix (Eq.~\ref{eq:master}) and substitute the associated off-diagonal matrix element into Eqs.~\ref{eq:chi} and ~\ref{eq:coeffdef}. Finally considering near-resonant operation, we expand the coefficient about $\delta = 0$ to arrive at,
\begin{eqnarray}
	\alpha^{(2)} = \frac{2\xi/\Gamma}{1 + 8\Omega_s^2/\Gamma^2} + O(\delta^2),
	\label{eq:alpha2}
\end{eqnarray}
where we have introduced a medium-dependent constant $\xi = 2\mathcal{N} d_{ab}^2\omega_s/(\hbar \epsilon_r \epsilon_0 n c)$, and noted that ${\rm Re}[\chi_{ab}] = 0$ and $n = 1$ at resonance. 

\begin{figure}[tb!]
\centering
\includegraphics[width=0.8\columnwidth,clip]{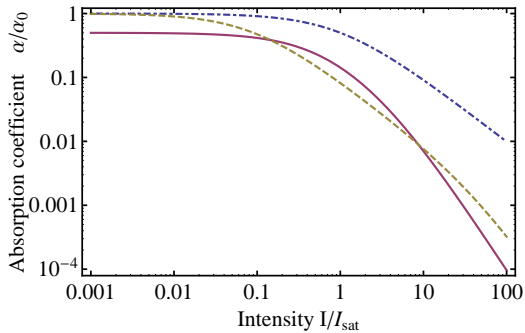}
\caption{Comparison of the absorption coefficient of the two-state and three-state systems. Dash-dot curve: two-state $\alpha^{(2)}$. Dashed: three-state $\alpha_s$ with $I_{\rm coh}/I_{\rm sat}=50$, $I_p/I_{\rm sat} = 0.1$. Solid: $\alpha_s$ with $I_{\rm coh} = I_p = I_{\rm sat}$. Beyond the saturation or coherent points, the scaling with the intensity is very different for the two systems with the the two-state (three-state) system being $\alpha/\alpha_0 \sim I^{-1}$  $(\sim I^{-2})$.}
\label{fig:alpha}
\end{figure}

We rewrite this expression as $\alpha^{(2)} = \alpha_0/(1 + I/I_{\rm sat})$, where $\alpha_0$ is the small signal or linear absorption coefficient, and the input field intensity $I = \zeta\Omega_s^2$ where $\zeta = \hbar^2 c\epsilon_0 \epsilon_r/(2d^2)$.
The medium containing multiple two-level system will become optically saturated when the signal intensity is strong, leading to reduced absorption. This turn-on of transmittance occurs at the saturation intensity $I_{\rm sat} = \zeta\Gamma^2/8$ when $\Omega_s$ exceeds the decay rate. This is shown by dash-dot curve in Fig.~\ref{fig:alpha}.
This form of scaling with $I$ in Eq.~\ref{eq:alpha2} is also found in three- and four-level schemes of saturable absorbers where a coherent driving of a transition with a second field is not considered~\cite{hercher67,sharan01}. By contrast, the ground-state coherence included in our model is responsible for an absorption scaling different from Eq.~\ref{eq:alpha2}.

\begin{figure}[tb!]
\centering
\includegraphics[width=0.8\columnwidth,clip]{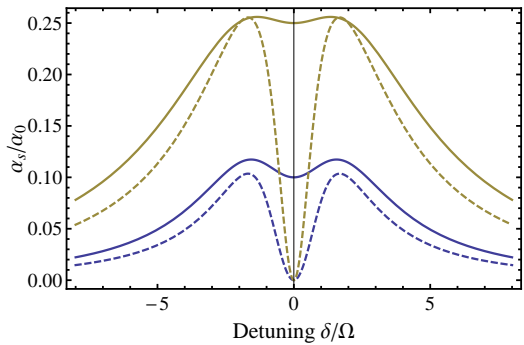}
\caption{Absorption spectra of the three-state system with different decay rate $\Gamma$ in the absence (dashed) and presence (solid) of dephasing ($\gamma = \Gamma/2$). We use $\Omega_p = \Omega_s = \Omega$, and $\Gamma/\Omega = 1$ (blue), 5 (yellow). When $\gamma = 0$, increasing $\Gamma$ effects the narrowing EIT window but this diminishes with $\gamma$.}
\label{fig:Gamma}
\end{figure}

We now turn to absorptive properties in three-state systems. The emergence of EIT can be understood in the dressed-state picture of the Hamiltonian Eq.~\ref{eq:Ham}~\cite{fleischhauer05}. At two-photon resonance ($\delta = 0$), the Hamiltonian has an eigenstate that is completely decoupled from the excited state: $|\mathcal{D}\rangle = (\Omega_s^2+\Omega_p^2)^{-1/2}(\Omega_p|a\rangle - \Omega_s|c\rangle)$. By being in state $|\mathcal{D}\rangle$ under the application of the two fields satisfying $\delta = 0$, system population in the dissipative state $|b\rangle$ is zero, so that there is no spontaneous emission and hence absorption of both fields cannot occur. This transparency window is positioned between two symmetrical absorption peaks (Autler-Townes (AT) splitting), which corresponds to transitions from $|\mathcal{D}\rangle$ to the other two dressed states. In the typical perturbative limit of $\Omega_s \ll \Omega_p$, the peak separation is $2\Omega_p$. This separation increases with $\Omega_s$, and we solve $d\alpha_s/d\delta = 0$ (with $\gamma, \Delta = 0$) to find that the peak maxima are centred at,
\begin{equation}
	\delta_{\pm}^0 = \pm \frac{(\Omega_p^2 + \Omega_s^2)^{3/4}}{\sqrt{\Omega_p}}.
	\label{eq:peakdelta0}
\end{equation}
where $\alpha_s$ denotes the absorption coefficient of the three-state scheme in the steady state. Notably while increasing spontaneous emission has the effect of broadening each peak (linearly as $2\Gamma$, when $\Gamma \gg \Omega_p, \Omega_s$) and narrowing the EIT window, the peak positions remain unchanged, as shown in Fig.~\ref{fig:Gamma}.

\begin{figure}[tb!]
\centering
\includegraphics[width=0.8\columnwidth,clip]{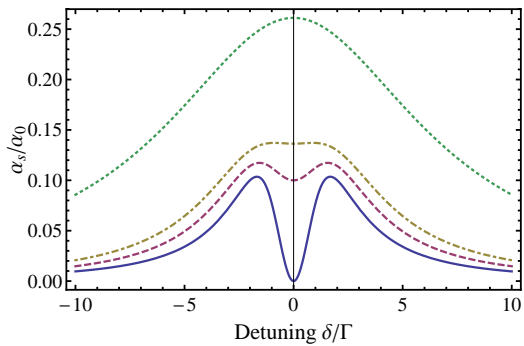}
\caption{Absorption spectra of the three-state system with increasing dephasing rate $\gamma/\Gamma = 0$ (solid), 0.5 (dashed), 1 (dash-dot), 5 (dotted). The other parameters are $\Omega_p = \Omega_s = \Gamma$.}
\label{fig:dephasing}
\end{figure}

In the presence of dephasing, field absorption between the two peaks becomes finite. While this has so far represented a fundamental limit for many of the EIT applications, this effect is exploited here. As the dephasing rate increases, the peak-to-peak distance of the AT splitting decreases and the overall absorption increases, as illustrated in Fig.~\ref{fig:dephasing}. Solving for the peak maxima yields,
\begin{equation}
	\delta_{\pm} = \pm \sqrt{ A\Big[ \frac{\Omega_p(5\gamma+2\Gamma)}{\Gamma'} + \frac{\Omega_s^2}{\Omega_p}\Big] - \frac{4\gamma A^2}{\Gamma'} }
	\label{eq:peakdelta}
\end{equation}
where $A = (\gamma\Gamma' + \Omega_p^2 + \Omega_s^2)^{1/2}$ and $\Gamma' = \gamma + 2\Gamma$ is the total decoherence rate. When the absorption peak is centred at zero detuning, Eq.~\ref{eq:peakdelta} becomes imaginary. In general provided that $\Omega_p^2, \Omega_s^2 \gg \gamma\Gamma$, all the important features of EIT remain observable. 

If we restrict our analysis to the case of single and near two-photon resonances with $\Delta = 0$ and $\delta \approx 0$, we can express the absorption of the signal field compactly under the condition $\dot{\rho} = 0$,
\begin{equation}
	\alpha_s = \frac{2\xi/\Gamma'}{1+\frac{\Omega_s^2}{\Omega_p^2} + \frac{12\Omega_s^2}{\Gamma\Gamma'} +\frac{(\Omega_p^2+\Omega_s^2)^2}{\Omega_p^2\gamma\Gamma'}} + O(\delta^2).
	\label{eq:alpha3}
\end{equation}
Note that because we have assumed near two-photon resonance, $\rho_{ab}$ is almost purely imaginary and the phase shift is near zero. Obviously this condition is strictly only true for the regime that the signal is a true continuous-wave (cw) field, but because we expect the non-zero absorption to always dominate the phase shifts, this approximation has little effect. 

For the simplified case of equal dipole moments $d_{ab} = d_{bc} = d$, the expression becomes
\begin{equation}
	\alpha_s = \frac{\alpha_0}{ 1 + I\Big(\frac{1}{I_p} + \frac{1}{I_{\rm sat}} + \frac{2}{I_{\rm coh}} \Big) + \frac{I_p}{I_{\rm coh}} + \frac{I^2}{I_p I_{\rm coh}} }
	\label{eq:alpha3int}
\end{equation}
with the intensity of the pump field $I_p = \zeta \Omega_p^2$ and now the saturation intensity is $I_{\rm sat} = \zeta\Gamma\Gamma'/12$. We can define $I_{\rm coh} = \zeta\gamma\Gamma'$ as the coherence intensity. Consequently $\Omega_p$ and $\Omega_s$ can only be varied through adjusting their respective field intensities.

It should be noted that this result is distinct from, but compatible with the previous works examining the special case of a single monochromatic field interacting resonantly with both optical transitions in the spectrally-broad pulse limit~\cite{koch86,khanin90} and the case of two input fields of equal intensity~\cite{koch90}. Specifically to retrieve the attenuation coefficient part of Eq.~3.5 in Ref.~\cite{koch90}, we equate $I_p = I$ to yield
\begin{equation}
	\alpha_s = \frac{\alpha_0}{ 1 + 3I/I_{\rm sat} + I/I_{\rm coh}},
	\label{eq:alpha3equal}
\end{equation}
and the saturation and coherence intensities are defined in the same way. Similar to the two-state systems, both of the systems in Eqs.~\ref{eq:alpha3int} and \ref{eq:alpha3equal} can become saturated when the signal field intensity is greater than saturation intensity. In addition, field absorption diminishes when $I > I_{\rm coh}$ in the effect known as coherent bleaching, which is different from saturation bleaching because it occurs even when the usual saturation condition is not met -- i.e. $I_{\rm sat} > I > I_{\rm coh}$~\cite{koch86,koch90}.

\begin{figure}[tb!]
\centering
\includegraphics[width=0.8\columnwidth,clip]{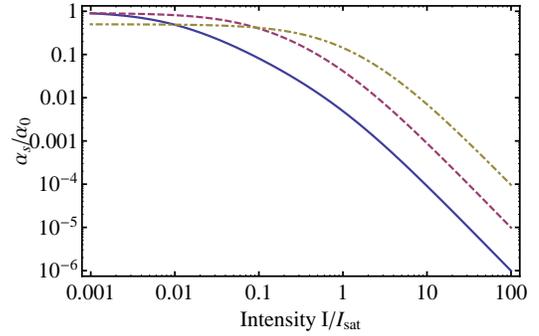}
\caption{Tuning the absorption coefficient of the three-state system with the pump field intensity $I_p$. The parameter values are $I_{\rm coh} = I_{\rm sat}$, and $I_p/I_{\rm sat}$ is 0.01 (solid), 0.1 (dashed), and 1 (dash-dot).}
\label{fig:tunablealpha}
\end{figure}

Closer inspection of Eq.~\ref{eq:alpha3int} reveals two additional circumstances under which the absorption of the three-state scheme changes with signal intensity. The first is when $I_{\rm sat}, I_{\rm coh}, \sqrt{I_p I_{\rm coh}} > I > I_p$. The coefficient turns from $\alpha \approx \alpha_0$ ($I < I_p$) to become inversely proportional to $I$, i.e., $I_p\alpha_0/I$, when the incident field exceeds $I_p$. Secondly, when the intensity exceeds $I_{\rm coh} + 2I_p$, the system shows a quadratic reduction in absorption with $\alpha \approx I_p I_{\rm coh} \alpha_0/I^2$. 
This is illustrated in Fig.~\ref{fig:alpha}.
This form of absorption nonlinearity represents an alternative to the conventional two-state saturable absorbers as a dynamically tunable nonlinear filter with a potentially greater intensity sensitivity. We refer to this effect as \emph{photon-assisted bleaching}. The bleaching point for rapid transmittance change can be adjusted by varying the pump field strength (see Fig.~\ref{fig:tunablealpha}) over a wide range $I_p \in (0, \min\{I_{\rm sat}, I_{\rm coh}\}]$ where saturation and coherent bleaching begin to dominate at the upper limit. Beyond this regime, we note that for
\begin{equation}
	I > \frac{I_p I_{\rm coh} + I_{\rm sat}I_{\rm coh} + 2I_p I_{\rm sat}}{I_{\rm sat}},
	\label{eq:quadcond}
\end{equation}
the coefficient changes $\sim C/I$ (for some constant $C$) to the quadratic scaling, depicted by the lower end of the dashed curve in Fig.~\ref{fig:alpha}.

Our analyses and observations are valid for a signal field with a small spectral width compared to the size of the EIT features at $\delta = 0$, because we have only used the zeroth-order term of $\alpha_s$. This bandwidth requirement fundamentally limits the operational speed of the three-state scheme. For instance, the FWHM of the two-state $\alpha^{(2)}$ is given by $\Delta\omega^{(2)} = \Gamma\sqrt{1+I/I_{\rm sat}}$ and the response time of the system $t_r$ commensurates with $2\pi/\Delta\omega^{(2)}$, which is necessarily faster than the excited-state lifetime. For the EIT-based scheme, we use the result in Eq.~\ref{eq:peakdelta} to estimate the allowed bandwidth $\Delta\omega \approx  (\delta_{+} - \delta_{-})/2$ as the distance between the peak maxima when the EIT window is still visible. We expand the width for $\gamma \ll \Gamma, \Omega_p, \Omega_s$ to find,
\begin{equation}
	\Delta\omega \approx |\delta_{\pm}^0| \Big\{ 1 + \frac{\gamma[\Gamma^2 + 2\Omega_p(\Omega_p - \sqrt{\Omega_p^2+\Omega_s^2})]}{2\Gamma(\Omega_p^2+\Omega_s^2)} \Big\},
\end{equation}
which increases as $\Omega_p\sqrt{1+I/I_p}$ to first order, and thus the bandwidth is tunable with the pump field. In the regime of dominant dephasing, we retrieve a single-peak absorption curve that simplifies to
\begin{equation}
	\alpha_s \approx 1/(1+4\delta^2/\Gamma'^2)
\end{equation}
with FWHM of $\Delta\omega = \Gamma'$. 

Our results and discussion so far provide the foundation of the proposed scheme for optically-thin absorber. However, to determine the overall response of an extended medium, we need to consider the evolution of the propagating fields. This is the subject of the next section.

\section{Optically-thick media}
\label{sec:opticallythick}
In an optically-thick system, the intensity of the transmitted fields must be treated as a function of position within the absorbing medium. Under the cw illumination or when the pulse has a broad and smooth distribution, the Beer-Lambert law can be used to determine the propagation of the pulse,
\begin{equation}
	\frac{d}{dz}I(z) = -\alpha(z)I(z),
	\label{eq:BLlaw}
\end{equation}
where $z$ is the depth in the medium. This can also be regarded as a short-memory Markov-like condition on the time evolution. There are at least two possible implementations, as depicted in Figs.~\ref{fig:schematics}(b) and (c), namely \emph{uniform-pump} and \emph{copropagating-pump} arrangements, respectively. We discuss the physics of each arrangement in details as follows.

\begin{figure}[tb!]
\centering
\includegraphics[width=0.8\columnwidth,clip]{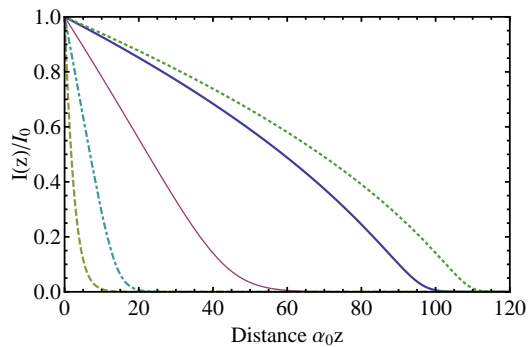}
\caption{Different decay rates of the intensity of the signal field as it propagates in a three-state medium in a uniform-pump arrangement. For all curves, $I_{\rm sat} = I_{\rm coh}$. Dotted curve: $2I_p = I_0/8 = I_{\rm sat}$. Bold: $I_p = I_0/10 = I_{\rm sat}$. Solid: $I_p/4=I_0/8 = I_{\rm sat}$. Dash-dot: $I_p = I_0/2 = I_{\rm sat}$. Dashed: $I_p = 5I_0 = I_{\rm sat}$. The first four curves show the tunability of the medium with the pump field.}
\label{fig:regimes}
\end{figure}

\subsection{Uniform-pump scheme}
In the uniform-pump arrangement, the pump field strength is uniform across the entire medium and only the signal field sees an optically-thick medium. The implementation therefore may involve directing the pump field at some angle to the propagation axis of the signal field, and thus this is most suitable in solid-state or cold atomic vapour environment where Doppler broadening is negligible. Since $I_p$ is a constant, we only need to integrate the propagation differential equation (Eq.~\ref{eq:BLlaw}) for the signal field using Eq.~\ref{eq:alpha3int} to arrive at the general law of intensity transfer for the three-state scheme,
\small
\begin{eqnarray}
	\alpha_0 z & = & \Big( 1 + \frac{I_p}{I_{\rm coh}} \Big)\ln\Big[\frac{I_0}{I(z)}\Big]  + \frac{I_0^2 - I(z)^2}{ I_p I_{\rm coh}} \nonumber \\
	& + & [I_0-I(z)]\Big( \frac{1}{I_p} + \frac{1}{I_{\rm sat}} + \frac{2}{I_{\rm coh}} \Big)
	\label{eq:transferlaw}
\end{eqnarray}
\normalsize
To solve for the characteristic penetration depth, we consider the different regimes separately. For the first forementioned case where bleaching occurs $I \gg I_p$, the penetration depth of the input field has extended from $L = 1/\alpha_0$ to $\sim I_0/(\alpha_0 I_p)$ at nonlinearity, comparable with the usual case of coherent bleaching (where $L \approx I_0/(\alpha_0 I_{\rm coh})$) and saturation ($L \approx I_0/(\alpha_0 I_{\rm sat})$). On the other hand in the second case, this length is given by $L \approx I_0^2/(\alpha_0 I_p I_{\rm coh})$.

If $I_0 \ll \max\{I_p,I_{\rm sat},I_{\rm coh}\}$, the absorptive nonlinearity is not manifested and the field decays exponentially in accordance with $I(z) \approx I_0\exp(-\alpha_0 z)$. However if $I_0 \gg \min\{I_p,I_{\rm sat},I_{\rm coh}\} = \bar{I}$, a linear decay occurs as $I(z) \approx I_0 - \alpha_0 \bar{I} z$ until the field intensity reaches $I(z) < \bar{I}$ that the decay becomes exponential. Finally when the condition in Eq.~\ref{eq:quadcond} is true, we find the slowest decay as $I(z) \approx I_0\sqrt{ 1- \alpha_0 I_p I_{\rm coh}z/I_0^2}$ that attenuates the signal field enough to become a linear decay, followed by an exponential drop off. These different bleaching behaviours are depicted in Fig.~\ref{fig:regimes} for variable pump field and incident field strengths.

\begin{figure}[tb!]
\centering
\includegraphics[width=0.8\columnwidth,clip]{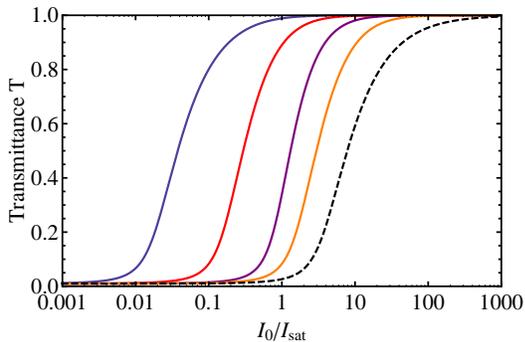}
\caption{(color online) Transmittance of an optically-thick (solid curves) three-state and (dashed) two-state media in an uniform-pump setup for different input signal strengths. For the three-state systems, we use $I_{\rm coh}/I_{\rm sat} = 1, T_0 = 0.01$ and variable pump field strength to adjust the bleaching level, in particular (from left to right) $I_p/I_{\rm sat} = 0.01$ (blue), 0.1 (red), 1 (violet), and 10 (orange). Observe that the turn on of the transmittance varies with pump field intensity and can occur with a sharper gradient than the two-state medium.}
\label{fig:transmittuni}
\end{figure}

To plot the transmittance of the medium of length $l$, we rewrite Eq.~\ref{eq:transferlaw} in terms of the small-signal filter transmission $T_0 = \exp(-\alpha_0 l)$ and overall transmittance $T = I(l)/I_0$, 
\small
\begin{equation}
	\ln\Big(\frac{T_0}{T}\Big) + I_0(1-T)\Big( \frac{1}{I_p} + \frac{1}{I_{\rm sat}} + \frac{2}{I_{\rm coh}} \Big) + (1-T^2)\frac{I_0^2}{I_pI_{\rm coh}} = 0.
\end{equation}
\normalsize
Treating $T_0$ as an independent parameter, we plot the solutions of this equation with $T_0 = 0.01$ (without loss of generality) in Fig.~\ref{fig:transmittuni} for different signal intensities. By varying the pump field strength, the transmission of the three-state-based filter is tunable and approaches unity when $I_0$ is greater than $I_p$. Importantly $T$ can approache unity at $I_0 \approx I_p$ more abruptly than in the two-level system because of the contribution of the quadratic term.

Group velocity reduction and spatial compression of the propagating fields occurs under the conditions of EIT. The group velocity of the signal field is related to the real part of the complex susceptibility (Eq.~\ref{eq:chi}),
\begin{equation}
	v_{g_s} = \frac{c}{n + \frac{\omega_s}{2}\frac{\partial Re[\chi_{ab}]}{\partial\delta} },
	\label{eq:vgs}
\end{equation}
where the parital derivative for $\Delta = 0, \delta = 0$ is
\small
\begin{equation}
	\frac{\partial Re[\chi_{ab}]}{\partial\delta} = \frac{2\mathcal{N} d^2 }{\hbar\epsilon_0\epsilon_r} \frac{\alpha_s}{\alpha_0} \frac{ 4 ( I_{\rm coh} + I )/\Gamma'^2 - ( I_p + I)/(\gamma\Gamma') }{ I_{\rm coh} + I + I_p  }.
	\label{eq:partialAB}
\end{equation}
\normalsize
Therefore the group velocity of the signal field can vary considerably depending on field intensities. In the regime of $I, I_p \ll I_{\rm sat}, I_{\rm coh}$, we find that
\begin{equation}
	v_{g_s} \approx \frac{c}{n + \frac{2\mathcal{N} d^2 \omega_s}{\hbar\epsilon_0\epsilon_r \Gamma'^2} }
	\label{eq:vgs_decoh}
\end{equation}
so that in the limit of fast spontaneous decay and/or dephasing the velocity reduction is minimized. On the other hand in the limit of small dephasing,
\begin{equation}
	v_{g_s} \approx \frac{c}{ n + \frac{2\mathcal{N} d^2 \omega_s}{\hbar\epsilon_0\epsilon_r \Omega_s^2} \frac{1}{(2 + I_p/I + I/I_p)} },
\end{equation}
where the velocity is also maximized with strong pump and/or signal field intensities. However the advantage of velocity reduction in EIT systems is that the medium interaction length $v_g t_r$ required for the atoms to reach the steady state can be reduced. 

\subsection{Copropagating-pump scheme}
In this setup, the signal and pump fields are copropagating and both see an optically-thick medium. While this arrangement can be implemented with solid-state systems, it is also suitable for media such as hot atomic gases, where one must work in a Doppler-free configuration to overcome the effect of Doppler broadening. Consequently, the absorption of the pump must be addressed and here we consider the absorption coefficient $\alpha_p$ of the pump field to investigate the field evolution within and transmittance of the medium in the similar way. By the symmetry of the system in the two-photon resonance regime, the associated coefficient has the same form as Eq.~\ref{eq:alpha3int} except $I_p$ is replaced with $I$ and vice versa,
\begin{equation}
	\alpha_p = \frac{\alpha_0}{ 1 + I_p \Big( \frac{1}{I} + \frac{1}{I_{\rm sat}} + \frac{2}{I_{\rm coh}} \Big) + \frac{I}{I_{\rm coh}} + \frac{I_p^2}{I I_{\rm coh}} }.
	\label{eq:alpha3pump}
\end{equation}
When one of the fields is completely absorbed, the medium becomes transparent to the other field. This is because, \textit{e.g.}, when the pump field is absorbed, the atoms are in the state $|c\rangle$ due to spontaneous decay and thus the absorption of the signal field can no longer occur. This happens when the ratio $\alpha_s/\alpha_p = I_p/I_0$ at entry to the system is less than unity, or $I_p < I_0$. As a result the pump field is attenuated at a much faster rate than the signal field and the values of the ratio and $\alpha_p$ increase with the distance. Conversely when $I_p > I_0$ and $\alpha_s/\alpha_p > 1$, the signal field is attenuated rapidly with increasing $\alpha_s$. Consequently the transmission response of the medium in this arrangement is more sensitive to the signal field intensities than the uniform-pump case. 

\begin{figure}[tb!]
\centering
\includegraphics[width=0.8\columnwidth]{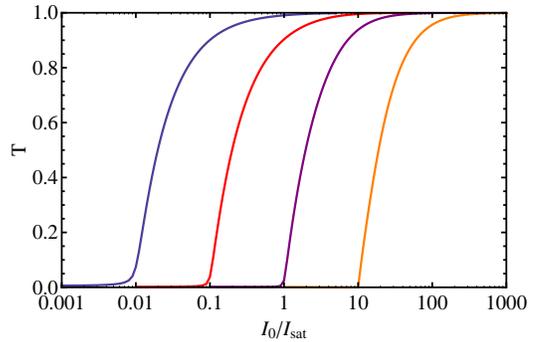}
\caption{Transmittance of an optically-thick three-state medium in the copropagating-pump arrangement with $I_{\rm sat} = I_{\rm coh}$. We vary the bleaching level with the pump field, specifically (from left to right) $(I_p/I_{\rm sat},\alpha_0 l) = (0.01, 5)$ (blue), (0.1, 7) (red), (1, 12) (violet), and (10, 70) (orange).}
\label{fig:transmittco}
\end{figure}

Under the assumption that the fields are travelling at the same group velocity, we solve Eq.~\ref{eq:BLlaw} as coupled differential equations with $\alpha_s$ and $\alpha_p$ in the same moving frame $z$ to plot $T$ versus $I_0$  in Fig.~\ref{fig:transmittco}. The sharp increase in transmittance of the medium to the signal field occurs at $I_0 = I_p$, and comparing with Fig.~\ref{fig:transmittuni}, this gradient is larger than the uniform-pump counterpart when the dephasing is weak.

Group velocity reduction under the conditions of EIT may diminish the observed properties of this copropagating arrangement. The mismatch of group velocity between the signal and pump fields can lead to reduction in effective two-beam interaction and different absorption behaviours akin to a two-state scheme. The solid-state based implementation can however take advantage of available techniques to engineer the group velocity of one of the fields to match the other, e.g., by tailoring the dispersion properties of the optical channel such as in photonic bandgap materials~\cite{krauss07} or coupled-resonator optical waveguides~\cite{altug05}. This control is not available in atomic vapour systems, but we note that there is a restricted operating conditions where the effect of mismatch can be minimized, as discussed below.

The group velocity $v_{g_p}$ of the pump field can be written down in an analogous way to Eqs.~\ref{eq:vgs} and \ref{eq:partialAB},
\begin{equation}
	v_{g_p} = c/\Big(n + \frac{\omega_p}{2}\frac{\partial Re[\chi_{bc}]}{\partial\Delta} \Big),
	\label{eq:vgp}
\end{equation}
where
\begin{equation}
	\frac{\partial Re[\chi_{bc}]}{\partial\Delta} = \frac{2\mathcal{N} d^2 }{\hbar\epsilon_0\epsilon_r} \frac{\alpha_p}{\alpha_0} \frac{ 4 ( I_{\rm coh} + 2I_p )/\Gamma'^2 }{ I_{\rm coh} + I + I_p  }
	\label{eq:partialBC}
\end{equation}
such that $v_{g_s}$ and $v_{g_p}$ are different in general. However in the same operating condition as Eq.~\ref{eq:vgs_decoh}, we find that $v_{g_p} \approx v_{g_s}$ if $\omega_s \approx \omega_p$, and the mismatch with the signal field velocity is minimized.

\subsection{Noise filtration}
\label{sec:noise}
Any absorbing material can be used to build a saturable absorber for signal-noise discrimination and a wide range of systems ranging from dyes, synthetic crystalline materials, to semiconductor quantum wells and dots are widely used~\cite{steinmeyer06}. Our EIT-based scheme may also be useful as a dynamically-tunable absorber for similar applications. As an illustration, we demonstrate noise filtering by simulating a transmission of a corrupted signal pulse through the three-state medium in the uniform-pump setup. In particular, we lift the restrictions for the use of the Beer-Lambert law and consider the Maxwell-Bloch (MB) equation to study the field propagation.

\begin{figure}[tb!]
\centering
\includegraphics[width=1\columnwidth]{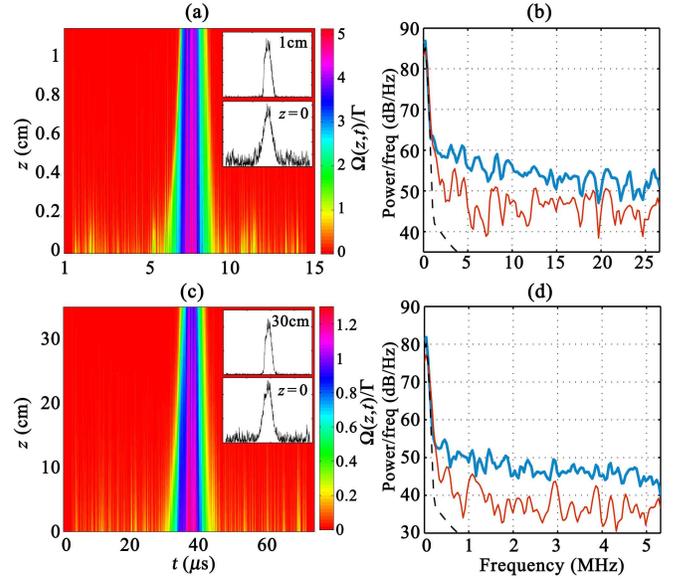}
\caption{(color online) Simulation of noise filtration of the signal field $\Omega(z,t)$ according to Eq.~\ref{eq:discreteMB} in two steps (a) $\Omega_p = \gamma = \Gamma = 10^7$~rad/s, $\zeta = 10^{11}$~m$^{-1}$s$^{-1}$, and (c) $\Omega_p = \Gamma/4$, $\gamma = \Gamma = 10^7$~rad/s, $\zeta = 10^9$~m$^{-1}$s$^{-1}$. The Gaussian signal pulse with pink noise is initialized at $z = 0$ and the noise components are absorbed as the signal propagates in the medium. The power spectral densities of the signals are shown in (b,d) respectively. Dashed curve: spectral density of the original signal. Bold blue: original corrupted signal. Solid red: filtered signal.}
\label{fig:noise_power}
\end{figure}

We begin by invoking the slowly-varying amplitude and phase envelope approximation to transform the second-order Maxwell's wave equation into a first-order MB equation for the real field amplitude,
\begin{equation}
	\Big( \frac{\partial}{\partial z} + \frac{1}{v_g} \frac{\partial}{\partial t}\Big) E_s(z,t) = \frac{\mathcal{N} d\omega_s}{2\epsilon_0\epsilon_r}{\rm Im}[\rho_{ab}].
\end{equation}
It is convenient to solve the propagation equation in terms of the Rabi frequency $\Omega_s(z,t)$ that reads
\begin{equation}
	\Big( \frac{\partial}{\partial z} + \frac{1}{v_g}\frac{\partial}{\partial t} \Big) \Omega_s(z,t) = -\frac{\alpha}{4} \Omega_s(z,t).
\end{equation}
As the medium polarization is affected by the field that is generates, we solve this equation by discretizing the expression,
\small
\begin{eqnarray}
 & & \frac{2}{\Delta z}\Omega_{j+1}^n + \frac{1}{2v_{g_{j+1}}^{~n}\Delta t}(\Omega_{j+1}^{n+1} - \Omega_{j+1}^{n-1}) + \frac{\alpha_{j+1}^n}{4}\Omega_{j+1}^n = \nonumber \\
 & & \frac{2}{\Delta z}\Omega_j^n - \frac{1}{2v_{g_j}^{~n}\Delta t}(\Omega_j^{n+1}-\Omega_j^{n-1}) - \frac{\alpha_j^n}{4}\Omega_j^n
 \label{eq:discreteMB}
\end{eqnarray}
\normalsize
where the subscript $s$ for $\Omega_s$ is dropped, $n = 1, 2, ..., N$ is the time step of size $\Delta t = T/N$ and $T$ is the simulation time. The index $j = 1, 2, ..., J$ is the sampling index for $z$ where a length unit in the medium of length $l$ is $\Delta z = l/J$. The expression can be recast in the matrix form ${\bf A}_{j+1}{\bf \Omega}_{j+1} = {\bf B}_{j}{\bf \Omega}_j$. At each step $j$ in space, the expression ${\bf \Omega}_{j+1} = [{\bf A}_{j+1}]^{-1} {\bf B}_j{\bf \Omega}_j$ is iteratively evaluated to reach convergence. The inverse matrix $[{\bf A}_{j+1}]^{-1}$ is calculated with initial trial values of ${\bf \Omega}_{j+1}$ given by the previous step ${\bf \Omega}_j$, and the matrices ${\bf A}_{j+1}$ and ${\bf \Omega}_{j+1}$ are solved for self-consistency.

In the simulation the atoms in the medium ($z > 0$) are initialized in the ground state $|a\rangle$ (i.e. the steady state when the signal field is switched off) and the signal field $\Omega(0,t)$ is a Gaussian pulse corrupted with $1/f$ pink noise components. The signal is initialized at entry to the medium for simplicity so that we do not treat the in-coupling of the field to the medium. The distribution and the spectral density of the input signal are shown in Fig.~\ref{fig:noise_power}. In Fig.~\ref{fig:noise_power}(a,b), we consider the case where the pump intensity is commensurate with the saturation and coherent bleaching intensities. In Fig.~\ref{fig:noise_power}(c,d), the case of relatively weaker signal and pump fields is simulated. In both cases, the high-frequency noise components are absorbed, showing the retrieval of the Gaussian signal. A temporally broad pulse is required for the second case because the system takes a much longer time to reach steady state to produce the absorptive behaviour predicted by Eq.~\ref{eq:alpha3}.

\section{Implementations}
\label{sec:impl}
The proposed scheme for tunable absorption nonlinearity can be implemented in a variety of solid-state and vapour phase media. We concentrate our attention on diamond defects for the solid-state and rubidium (Rb) vapour for the gas phase, but we note in passing that quantum dots~\cite{chang06,xu08}, and rare-earth crystals~\cite{ham97} provide equally promising solid-state implementations.

In the solid state, we envisage one practical realization using a doped diamond waveguide implemented as a step-index fibre~\cite{olivero05,hiscocks08}, a line defect in the photonic-bandgap structure~\cite{notomi08,krauss07} [Fig.~\ref{fig:schematics}(a)], or in coupled-resonator configuration~\cite{notomi04,altug05}. Diamond hosts numerous optically-active impurity- and/or vacancy-based defects, and a prominent centre featured in numerous quantum-optical device designs is the negatively-charged nitrogen-vacancy (NV) centre~\cite{greentree08}. Coherent population trapping due to the same mechanism responsible for EIT has been demonstrated with the centre~\cite{santori06,santori06b}. 

By operating near the zero-phonon line resonance of the centre $\lambda = 638$~nm, the NV can be treated as a $\Lambda$-type system where we identify the required energy levels with the available spin states, $|b\rangle = |^3 E,m=0\rangle$ being the excited spin triplet state whereas states $|a\rangle$ and $|c\rangle$ correspond to the $m=0$ and $m=+1$ (or $m=-1$) Zeeman sublevels of the triplet ground state $|^3 A\rangle$~\cite{manson06}. The dipole moment between transitions $|a\rangle-|b\rangle$ and $|c\rangle-|b\rangle$ is $\sim 10^{-30}$~Cm~\cite{su09}. To avoid optical cross coupling between the NVs, they should be spatially separated by $\geq \lambda/n \approx 250$~nm in the single-mode waveguides with a cross-section area of $200\times 200$~nm$^2$, limiting the atom number density to $\mathcal{N} \leq 10^{20}$~m$^{-3}$. This density level can also controlled and achieved using ion implantation in synthetic diamond~\cite{meijer05}. Consequently, with $\epsilon_r = 10$, the characteristic constant $\xi = 7\times 10^{10}$~m$^{-1}$s$^{-1}$. 

The NV has a radiative lifetime of 11.6~ns or $\Gamma/\pi = 86$~MHz and a dephasing lifetime up to $1~\mu$s~\cite{dutt07}, so that the small-signal attenuation coefficient is $\alpha_0 = 2\xi/\Gamma' = 244$~m$^{-1}$ and saturation and coherence intensities are of order 1~MW/m$^2$. For $T_0 = 0.01$ in Fig.~\ref{fig:transmittuni}, the required length $l$ of medium to demonstrate the tunable absorption properties for the uniform-pump setup is 2~cm. Similarly for the copropagating-pump implementation in Fig.~\ref{fig:transmittco}, we estimate $l = 2$~cm (for $I_p/I_{\rm sat} = 0.01$) and an interaction length of 29~cm (for 10) for the same value of absorption coefficient. This can be achieved by folding the waveguide back onto itself for multiple times in a standard diamond wafer. The dephasing time is related to the concentration of electron paramagnetic impurities in the diamond lattice~\cite{kennedy03} and hence $\alpha_0$ can be in part controlled at fabrication. 

The $\Lambda$-type configuration also appears naturally in the hyperfine structures of the atomic $^{87}$Rb vapour. Using the $D_1$ transition, the pump and signal fields of wavelength $\lambda = 795$~nm couple pairs of Zeeman sublevels of the atomic $5 S_{1/2}$, $F=2$ state via the common upper state $5 P_{1/2}$, $F=1$~\cite{li96,figueroa06}. The overall rate of spontaneous emission into the ground states is $\Gamma/\pi = 37$~MHz so that the dipole moment is $d \sim 10^{-29}$~Cm and the saturation intensity is $\sim 350$~W/m$^2$. The usual condition of weak ground-state dephasing of $\gamma/2\pi = 117$~Hz~\cite{figueroa06} in an typical EIT experiment is not required in our scheme because the coherent bleaching would become the dominant effect. The atomic density can vary from $10^{10}$ to $10^{15}$~cm$^{-3}$ in a typical 5~cm cell at room temperature~\cite{li96,phillips01}. Suppose that a density level $\mathcal{N} = 10^{21}$~m$^{-3}$ and $\xi = 5\times 10^{14}$~m$^{-1}$s$^{-1}$. In a velocity-matched, copropagating-pump or Doppler-free uniform pump setups, we expect a short cell of length of $10 - 100$~$\mu$m should be sufficient for operations with $I_p/I_{\rm sat} = 0.01$ and 0.1, respectively, owing to a relatively large coefficient $\alpha_0 \approx 2\times 10^6$~m$^{-1}$. 

\section{Conclusion}
We have analysed nonlinear absorption in an ensemble of three-level atoms in a signal-pump EIT configuration. In the presence of ground-state dephasing, EIT is not complete at two-photon resonance and bleaching occurs. While this is comparable with conventional saturable absorption in an ensemble of two-level atoms, the effective absorptive nonlinearity associated with the signal field in the EIT system can be stronger, and can scale quadratically with signal intensity. Moreover as the EIT features depend on the pump beam intensity, the bleaching intensity of the absorber can be modified post-fabrication. We considered two different signal-pump arrangements with uniform pump and copropagating pump beams, and show that nonlinear change in the transmission property of an optically-thick three-state medium is sharper than the two-state scheme. This effect may be useful for noise filtration and microscopy, especially in the weak signal regime, and perhaps in the future, for manipulating the quantum, rather than classical, properties of fields.

\section*{Acknowledgments}
We thank A. M. Martin, J. Q. Quach, and A. Hayward for helpful discussions. This project is supported by the Australian Research Council under the Discovery Scheme (DP0880466 and DP0770715), the Centre of the Excellence Scheme (CE0348250), and in part by MEXT and FIRST in Japan. CHS acknowledges the support of the Albert Shimmins Memorial Fund.

\end{document}